# GPT-4's One-Dimensional Mapping of Morality: How the Accuracy of Country-Estimates Depends on Moral Domain

**Pontus Strimling[1,2,3], Joel Krueger[1], Simon Karlsson[1]**

[1]The Institute for Future Studies, Stockholm, Sweden
[2]Department of Women's and Children's Health, Uppsala University, Uppsala, Sweden
[3]Institute for Analytical Sociology, Linköping University, Norrköping, Sweden

*Abstract*. Prior research demonstrates that Open AI's GPT models can predict variations in moral opinions between countries but that the accuracy tends to be substantially higher among high-income countries compared to low-income ones. This study aims to replicate previous findings and advance the research by examining how accuracy varies with different types of moral questions. Using responses from the World Value Survey and the European Value Study, covering 18 moral issues across 63 countries, we calculated country-level mean scores for each moral issue and compared them with GPT-4's predictions. Confirming previous findings, our results show that GPT-4 has greater predictive success in high-income than in low-income countries. However, our factor analysis reveals that GPT-4 bases its predictions primarily on a single dimension, presumably reflecting countries' degree of conservatism/liberalism. Conversely, the real-world moral landscape appears to be two-dimensional, differentiating between personal-sexual and violent-dishonest issues. When moral issues are categorized based on their moral domain, GPT-4's predictions are found to be remarkably accurate in the personal-sexual domain, across both high-income (*r* = .77) and low-income (*r* = .58) countries. Yet the predictive accuracy significantly drops in the violent-dishonest domain for both high-income (r = .30) and low-income (r = -.16) countries, indicating that GPT-4's one-dimensional world-view does not fully capture the complexity of the moral landscape. In sum, this study underscores the importance of not only considering country-specific characteristics to understand GPT-4's moral understanding, but also the characteristics of the moral issues at hand.

## 1 Introduction

Large Language Models (LLMs) form their understanding and knowledge through the processing of their pre-training data. Similar to humans, these models are susceptible to various biases, including those related to race, gender, political ideology, and religious affiliation (Abid et al., 2021; Kirk et al., 2021; Rozado, 2023; Singh & Ramakrishnan, 2023). While multiple efforts of bias-reduction have proven effective, such as the utilization of reinforcement learning with human feedback and the fine-tuning of training datasets, LLMs are still not immune to biases (Ray, 2023). As such, it remains an ongoing endeavor to understand where these models excel and fall short.

In this study, we focus on examining GPT-4's grasp of cultural nuances within the domain of moral opinions. The interest in studying moral opinions is twofold. Firstly, as people increasingly rely on LLMs like GPT-4 for information, it becomes crucial to understand their accuracy and potential biases across various domains, including that of morality. Our perceptions about other cultures influence how we interact with them in an increasingly globalized and digitalized world. Secondly, LLM's cultural knowledge may be instrumental for other tasks they perform. For example, one area gaining traction is the cultural and social alignment of AI agents (AlKhamissi et al., 2024; Lee et al., 2024; Lin & Chen, 2023). Given the global deployment of AIs, it becomes central to determine the types of values these agents should embody and to mitigate any unintended cultural biases. Previous research has proposed various strategies for enhancing cultural alignment between users and LLMs (AlKhamissi et al., 2024; Tao et al., 2023), such as country-specific prompting (e.g., instructing the LLM to answer like a person from another society). However, the efficacy of such approaches, and cultural alignment in general, is likely to depend on the accuracy of the LLM's cultural understanding, warranting a comparison between AI-generated output and real-world data on moral issues.

Prior research has established that GPT models contain some understanding of human moral standards (Almeida et al., 2024; Dillion et al., 2023; Schramowski et al., 2022). In a comprehensive study, Ramezani & Xu (2023) conducted a cross-cultural examination using GPT-2 and GPT-3. Their research demonstrated that both models' estimates of moral opinion were significantly correlated with cross-cultural variation in real-world survey data, but that GPT-3 exhibited superior performance compared to GPT-2. Moreover, they separated countries by their economic status and continent, revealing more accurate estimates in more affluent and Western countries. The authors concluded that the discrepancy in the accuracy of moral opinion representation between affluent, Western nations and those that are not affluent or non-Western raises concerns about the potential limitations of English-language training data. They further speculated that the observed gap may stem from either a deficiency in the representation of diverse cultural moral norms or the predominance of Western viewpoints in descriptions of other countries and cultures.

While previous studies have focused on investigating differences in GPT's accuracy between different categories of countries (Ramezani & Xu, 2023), limited attention has been paid to evaluating its accuracy across different types of moral issues. This despite an extensive literature advocating for multi-dimensional taxonomies of cultural values to comprehend cultural diversity (Graham et al., 2013; Haidt, 2007; R. F. Inglehart, 2018; R. Inglehart & Welzel, 2010; Schwartz, 2004; Vauclair & Fischer, 2011). Of particular interest is the distinction first made by Vauclair and Fischer (2011), proposing the separation of moral issues into two categories: personal-sexual and dishonest-illegal issues. Drawing from evolutionary and cultural theories, they hypothesized that the dishonest-illegal moral domain should be independent of cultural values, whereas personal-sexual issues should be influenced by cultural perceptions of the self (i.e., independent versus interdependent selves). This notion found support when examining the relationship between Shwartz (2004) cultural values and the two moral domains using World Value Survey data, where the Autonomy-Embeddedness dimension successfully predicted countries attitudes on personal-sexual but not violent-dishonest issues. Taken together, this example underscores the complexity of the moral landscape and emphasizes the importance of recognizing the diversity of moral issues.

In this study, we aim to get a better understanding of GPT-4's ability to predict variation in moral opinions by accounting for both characteristics of countries and the moral issues at hand. We begin by replicating Ramexani and Xu's (2023) finding that GPT contains knowledge about cultural variation in moral opinion and that it is significantly better at inferring moral opinions in high-income than in low-income countries. We then turn to the complexity of moral issues, comparing the factor structure of GPT estimates with real-world survey data. Finally, building upon the results of the factor analysis, we conduct a series of supplementary analyses to delve deeper into potential underlying causes and ramifications of GPT's dimensional structure and how it estimates moral opinions.

## 2 Material and methods

*EVS/WVS*

Like Ramezani and Xu (2023), we used the joint EVS/WVS 2017-2022 as our reference dataset. The dataset is a publicly available compilation of two large cross-cultural surveys, the World Value Survey Wave 7 (WVS) and the European Value Study 2017 (WVS/EVS, 2020). We used all items in the WVS Ethical and Values section that were included in both the WVS and the EVS, resulting in 18 moral items surveyed across 63 geographically, culturally, and economically varied countries. For each item, participants in the WVS/EVS were asked to judge how justifiable a behavior is (e.g. abortion) on a Likert scale ranging from 1 ("never") to 10 ("always"). We averaged participants' scores within countries to obtain country-level scores for each item. Henceforth, we will refer to the joint WVS/EVS dataset as simply WVS.

*ChatGPT*

We asked chatGPT 4 (October, 2023) to make country-level estimates for each of the 18 moral items in each of the 63 countries for which there is survey data. In the prompt, we specified the scale ranging from 1 ("never justifiable") to 10 ("always justifiable"), while requesting one decimal precision.

**Analysis**

*Accuracy*

To test GPT's accuracy, we calculated Pearson correlation coefficients between WVS scores and GPT estimates. To avoid inflated correlations due to average differences in justifiability between issues, scores were normalized across countries for each moral issue in the WVS data and the GPT estimates.

*Dimensionality*

We performed an exploratory factor analysis separately on the WVS data and the GPT estimates.

## 3 Evaluation and results

We evaluate the overall accuracy of GPT estimates, and also how accuracy differs between high-income and low-income countries. We then compare the dimensionality of the WVS data with that of GPT's estimates and investigate differences in GPT's accuracy between different kinds of moral issues.

*Overall accuracy*
As a first test of GPT-4's knowledge of moral cultural variation, we compared GPT estimates and WVS scores across all countries and issues in the WVS dataset. The results show WVS scores and GPT estimates to be moderately positively correlated, *r*(1114) = .47, *p* < 0.001, 95% CI [0.43, 0.52], confirming that the GPT model contains some information about societal variance in moral opinions. The mean correlations between GPT estimates and WVS scores for each moral issue separately revealed a mean correlation of .48, 95%, CI [0.33, 0.63], ranging from 0.03 for the justifiability of avoiding a fare on public transport to 0.90 for the justifiability of homosexuality. Further, splitting countries by income level revealed a significant difference between high-income and low-income countries, with a strong correlation in the high-income subset, *r*(748) = .50, *p* < 0.001, 95% CI [0.45, 55], but only a weak correlation in the low-income subset, *r*(364) = 0.16, *p* = 0.003, 95% CI [0.05, 0.25]. See Figure 1.

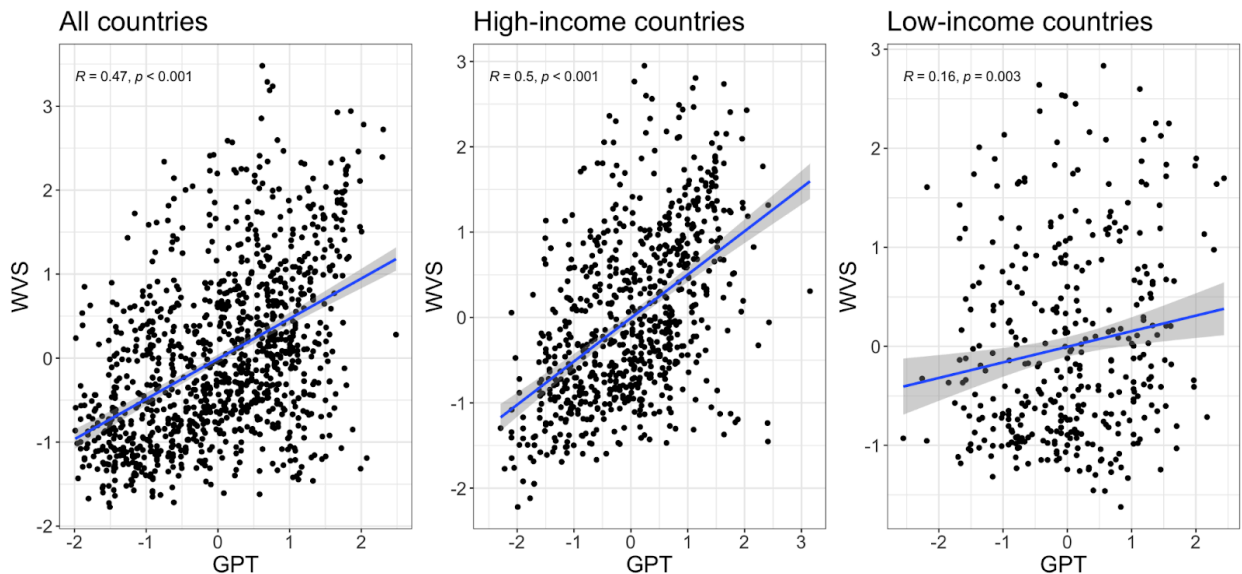

**Fig 1. Correlation between GPT estimates and WVS scores.** Each point represents public opinion in a country (*n* = 63) on a moral issue (*n* = 18) as estimated by GPT (x-axis) and recorded by the WVS (y-axis). The cut-off for high-income (*n* = 42) vs. high-income countries (*n* = 21) is a gross national income (GNI) per capita of $4,255. The shaded areas delineate 95% confidence intervals around the regression line.

*Dimensionality of WVS vs. GPT data*
To compare the dimensionality of the two datasets, we performed an exploratory factor analysis separately for the WVS and GPT data. A scree plot of eigenvalues revealed that two components should be retained for the WVS data and only one component for the GPT data (see Figure 2). The first two components for the WVS data had a cumulative explained variance of 0.78, which was identical to the explained variance of the first component for the GPT data (see bottom of Table 1).

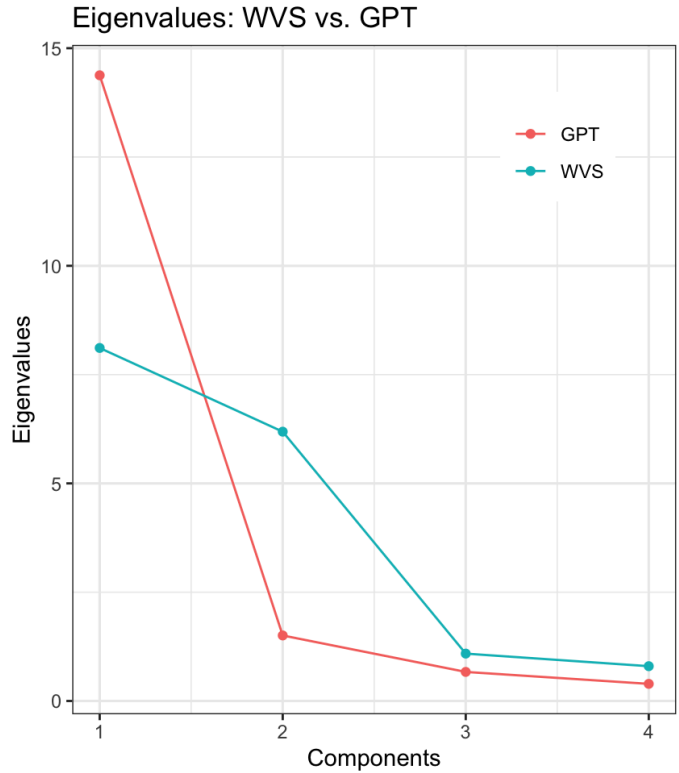

**Fig 2. Scree plot of eigenvalues for GPT vs. WVS.**

A factor analysis with varimax rotation retaining two factors for the WVS data showed that almost all of the 18 moral issues could be separated into two factors, with the exceptions of two issues with slight cross-loadings, the justifiability of suicide and parents beating their children, and one issue that did not load well on either factor, the justifiability of the death penalty. The loadings of the single factor for the GPT data support a similar split, with positive loadings for issues that correspond to the WVS first factor loadings, and negative loadings for issues that mostly correspond to the WVS second factor loadings. See Table 1.

Table 1.

*Factor loadings with varimax rotation, calculated separately for WVS and GPT*

| Moral items | WVS | | GPT |
| --- | --- | --- | --- |
| | Factor 1 | Factor 2 | Factor 1 |
| Abortion | **0.92** | 0.02 | **0.94** |
| Divorce | **0.91** | -0.12 | **0.97** |
| Euthanasia | **0.88** | 0.01 | **0.98** |
| Homosexuality | **0.95** | -0.06 | **0.94** |
| Casual sex | **0.94** | 0.12 | **0.94** |
| Prostitution | **0.94** | 0.17 | **0.94** |
| Sex before marriage | **0.95** | -0.04 | **0.94** |
| Suicide | **0.86** | 0.34 | **0.93** |
| Man beating wife | -0.13 | **0.94** | -0.97 |
| Violence | 0.16 | **0.95** | -0.82 |
| Stealing property | 0.15 | **0.97** | -0.83 |
| Claiming benefits | -0.06 | **0.69** | -0.77 |
| Avoiding fare | 0.16 | **0.74** | -0.73 |
| Cheating on taxes | 0.18 | **0.88** | -0.81 |
| Accept bribe | 0.11 | **0.97** | -0.85 |
| Death penalty | 0.22 | 0.10 | -0.71 |
| Political violence | 0.24 | **0.94** | -0.78 |
| Parents beating children | **-0.32** | **0.51** | -0.95 |
| % of variance | 0.40 | 0.38 | 0.78 |

*\* Loadings larger than 0.30 in bold*

*Personal-sexual vs. violent-dishonest issues*

Based on the GPT factor's positive versus negative loadings, we separated the 18 moral issues into two groupings or moral domains. Largely following the distinction made by Vauclair & Fischer (2011), we refer to the first category as *personal-sexual* issues (abortion, divorce, euthanasia, homosexuality, casual sex, prostitution, sex before marriage, and suicide) and to the second category as *violent-dishonest* issues (man beating wife, violence, stealing property, claiming government benefits to which one is not entitled, avoiding a fare on public transport, cheating on taxes, someone accepting a bribe in the course of their duties, the death penalty, political violence, and parents beating their children).

Given that GPT seems to derive all its predictions based on a single underlying factor, we would expect a high correlation between GPT's estimates in the two moral domains. On the other hand, we would expect little or no correlation between the same domains in real-world survey data, as the analysis suggested that two distinct factors drive variation in the WVS data. To test whether this notion finds support, we calculated country-level means for personal-sexual and violent-dishonest issues and examined the correlation between the two domains for GPT and WVS, respectively. As for the results, there was indeed a large correlation between GPT's country estimates in the two domains, $r(61) = -.86$, $p < 0.001$, 95% CI[-0.91, -0.78], while only a small, non-significant correlation was found in real-world data, r(52) = .14, *p* = .3, 95% CI[-0.13, 0.40]. See Figure 3.

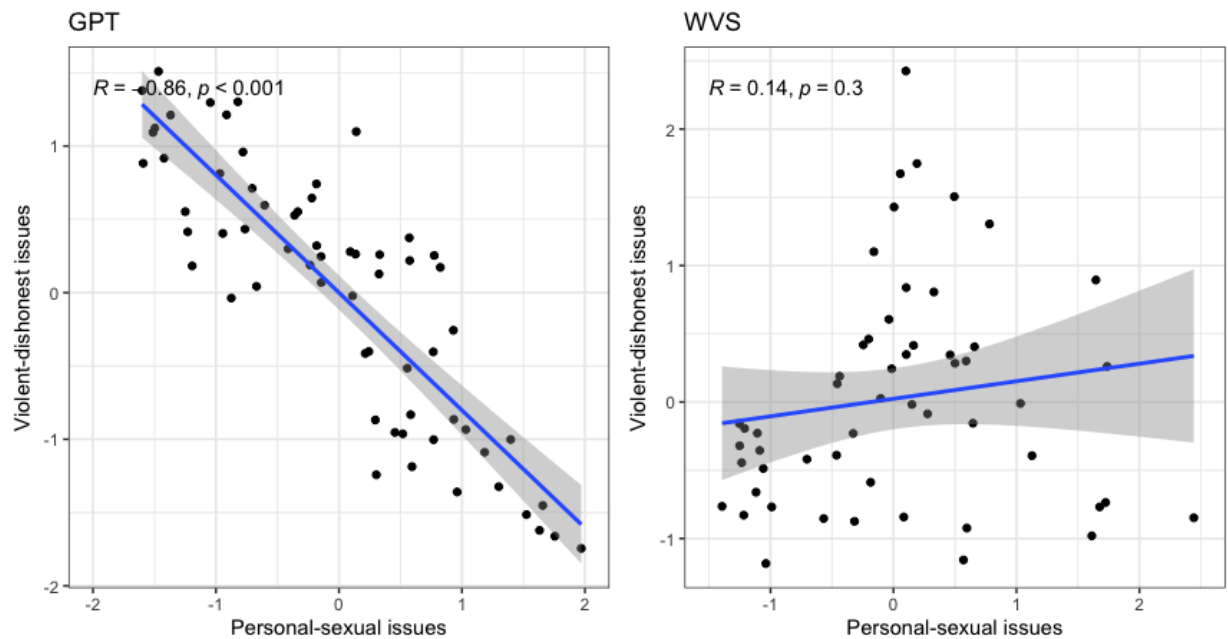

**Figure 3. Relationship between violent-dishonest and personal-sexual Issues: GPT vs. WVS.** Each data point represents the mean-level public opinion in a country for violent-dishonest issues (y-axis) and personal-sexual issues (x-axis), as estimated by GPT (left figure) or recorded by the WVS (right figure). The shaded areas delineate 95% confidence intervals around the regression line.

So far, we have shown that GPT-4 performs significantly better at estimating cultural variation across high-income countries than across low-income countries. We also found that GPT's estimations are primarily based on a single underlying factor, despite the analysis of real-world data suggesting a more nuanced explanation involving two factors. Together, these results suggest that it may not only be important to consider countries' income level when evaluating GPT's predictive success, but also the moral domain in which GPT makes its predictions. To account for both these factors, we tested accuracy for all four combinations of income and domain. The results, displayed in Figure 4, reveal that GPT not only performs well in the personal-sexual domain in high-income countries, *r*(331) = .77, *p* < 0.001, 95% CI [0.72, 0.81], but also in low-income countries *r*(156) = .58, *p* < 0.001, 95% CI [0.47, 0.68]. In contrast, GPT performs relatively poorly across both income groups in the violent-dishonest domain, with a weak to moderate correlation amongst high-income countries, *r*(415) = .30, *p* < 0.001, 95% CI [0.21, 0.38], and a weak *negative* correlation in low-income countries, *r*(206) = -0.16, *p* < 0.02, 95% CI [-0.29, -0.03]. GPT's high performance for personal-sexual issues and low performance for violent-dishonest issues, across income levels, suggests that the accuracy of estimates depends more on differences between different types of issues than on country income level.

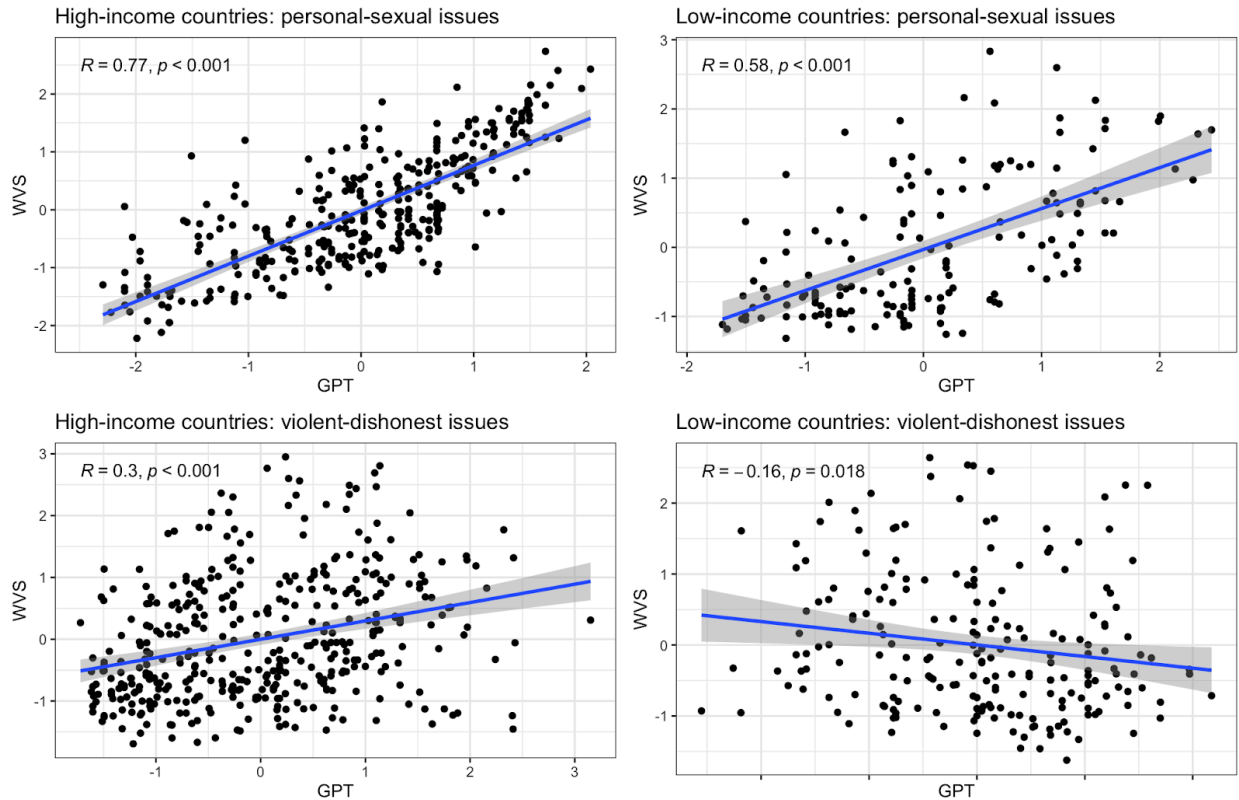

**Figure 4. GPT accuracy separated by country income level and moral domain.** Each point represents public opinion in a country (high- or low-income) for an issue (personal-sexual or violent-dishonest), as estimated by GPT (x-axis) and measured by the WVS (y-axis). The shaded areas delineate 95% confidence intervals around the regression line.

To get a better overview of the difference in accuracy between the two domains, we calculated country-level means separately for each domain and then correlated the GPT-means with the WVS-means. We found that GPT very accurately estimated societal variation in the personal-sexual domain, $r(61) = .85$, $p < 0.001$, 95% CI [0.76, 0.90], but performed poorly in the violent-dishonest domain, $r(61) = .23$, $p = 0.07$, 95% CI [-0.02, 0.45]. The right panel of Figure 5 clearly shows that GPT's single factor does not capture societal variation in the violent-dishonest domain. For instance, GPT's point estimates for Armenia and the Philippines are barely differentiated along the x-axis, though they are furthest apart in the empirical ratings recorded by the WVS, that is, along the y-axis.

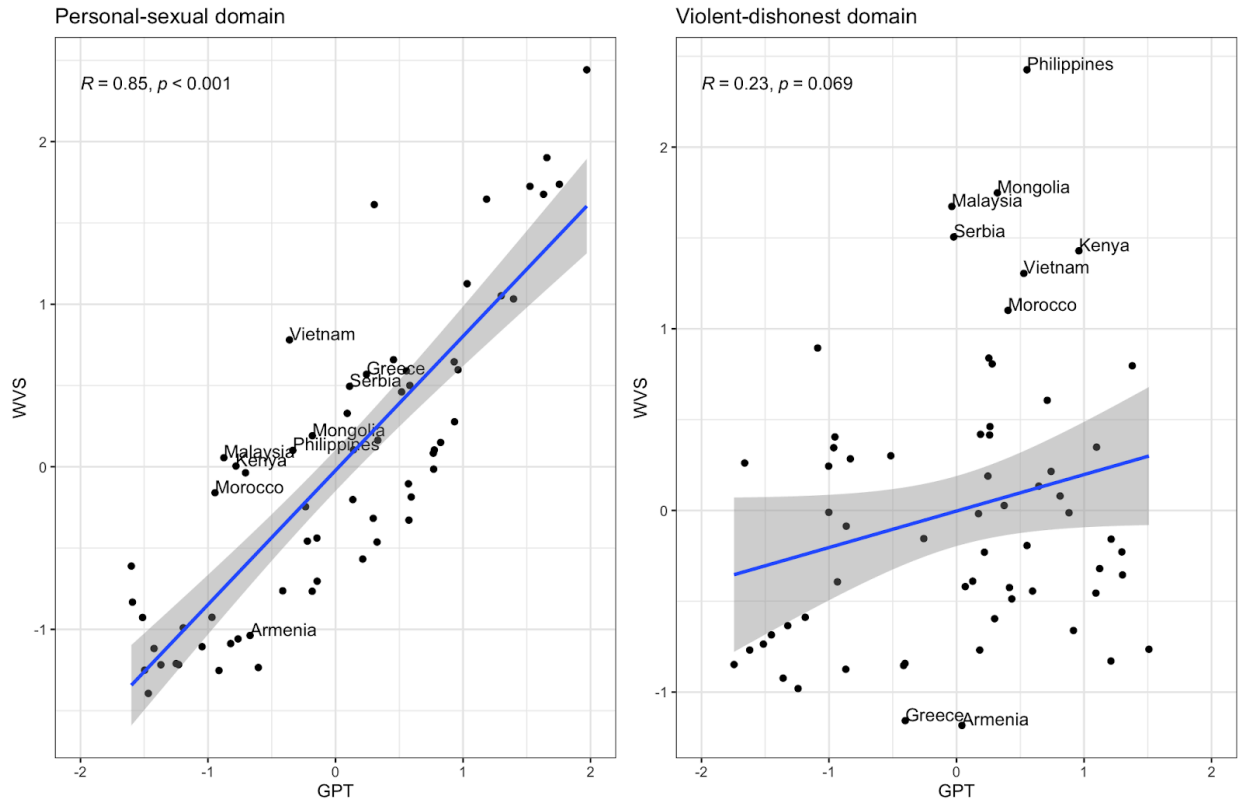

**Figure 5. Country-level means for personal-sexual and violent-dishonest moral issues.** Each point represents country-level means for personal-sexual moral issues (left panel) or for violent-dishonest issues (right panel), as estimated by chatGPT (x-axis) and measured by the WVS (y-axis). The shaded areas delineate 95% confidence intervals around the regression line.

Finally, we turned to the reason why GPT was so accurate in the personal-sexual domain while performing so poorly in the violence-dishonest domain. The issues in the personal-sexual domain, for which GPT is highly accurate, pertain to social topics that often divide liberals and conservatives. For this reason, we hypothesized that GPT's one-dimensional view of morality may mainly place countries along a liberal-conservative continuum. To explore this hypothesis, we asked GPT to estimate how "liberal/progressive" or "traditional/conservative" each of the 63 countries in the WVS dataset is, on a scale from 0 (conservative) to 100 (extremely liberal). We then correlated these liberalism estimates with the same country-level means used in the analysis above, that is, means across all issues in each domain for GPT estimates and WVS scores, respectively.

The results revealed a very strong positive correlation between GPT's liberalism estimates and GPT's mean country-level estimates in the personal-sexual domain, $r(61) = .93$, 95% CI [0.89, 0.96], as well as between liberalism estimates and the WVS country-level mean scores in the personal-sexual domain, $r(61) = .84$, 95% CI [0.75, 0.90]. This suggests that a liberalism-conservative dimension indeed appears to explain GPT's estimates for personal-sexual issues, and also that this dimension is sufficient to capture most of the real-world country variance in this domain of moral opinion. In contrast, there was a very strong

*negative* correlation between GPT's liberalism estimates and GPT's mean country-level estimates in the violent-dishonest domain, $r(61) = r(61) = -.85$, 95% CI [-0.91, -0.77], and no significant correlation between liberalism estimates and the WVS country-level mean scores in the violent dishonest domain, $r(61) = -.13$, 95% CI [-0.36, 0.12]. This suggests that a liberal-conservatism dimension also appears to explain GPT's estimates in the violent-dishonest domain, but that this continuum explains little to no real-world variance in this domain of moral opinion. See Figure 6.

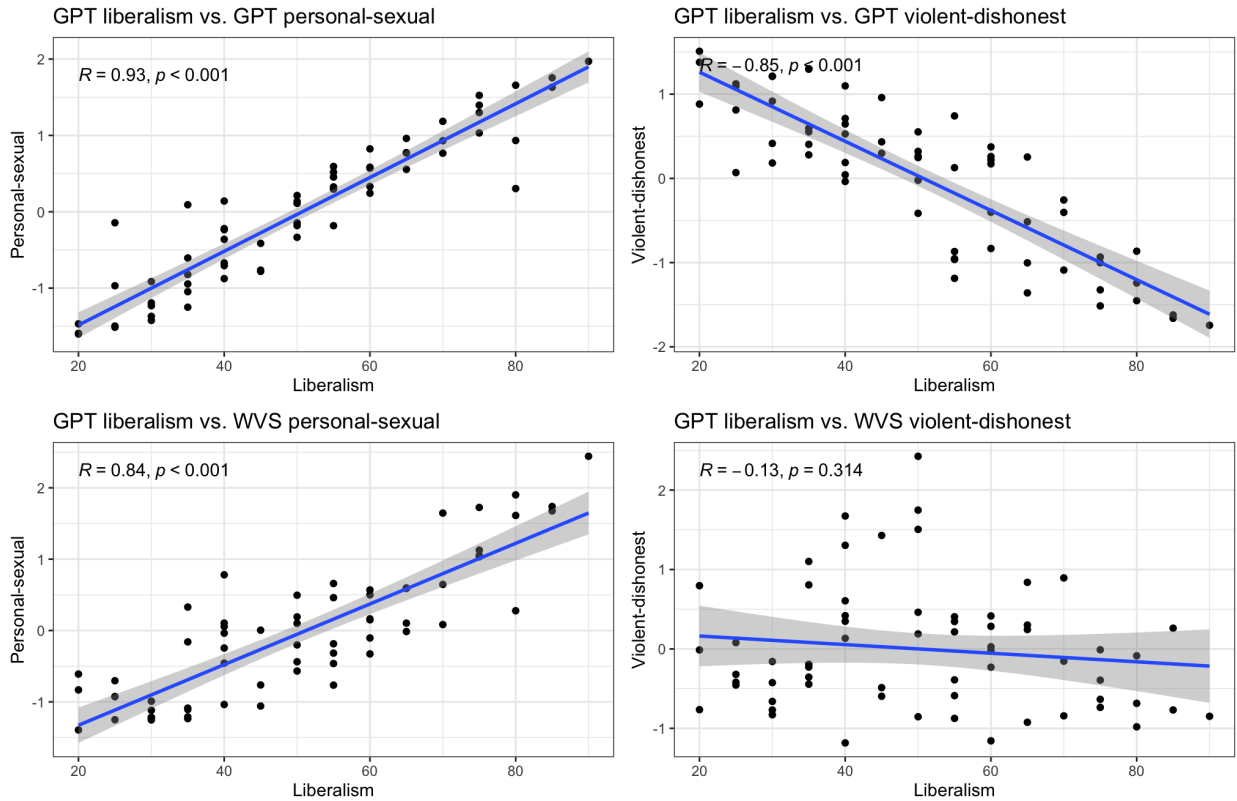

**Figure 6. Liberalism vs. GPT estimates and WVS scores in each moral domain.** The x-axis represents GPT's country estimates of liberalism (0 = conservative, 100 = extremely liberal). The y-axes represent GPT's country-level estimates and the WVS scores for the personal-sexual and violent-dishonest issues, respectively. The shaded areas delineate 95% confidence intervals around the regression line.

## 4 Discussion

In this study, we investigated whether GPT-4 could accurately estimate moral opinions across a large set of geographically, culturally, and economically varied countries. Replicating previous findings by Ramezani and Xu (2023), our first analyses showed that GPT's estimates do correlate with real-world data, with estimates being significantly more accurate in high-income than in low-income countries. Importantly, GPT estimates and real-world scores were not perfectly correlated for any moral issues, assuaging any worries that GPT may be basing its estimates on the WVS data. On their own, these results may be interpreted as mainly supporting a Western-centric or English-language bias in the model. However, closer scrutiny revealed a more complicated picture.

First, we found that a single factor could explain almost all variance in GPT's estimates of moral opinions, while two factors were needed to explain as much variance in the WVS survey data. In other words, GPT's moral worldview is almost entirely one-dimensional, while the actual moral landscape is largely two-dimensional. Second, factor loadings supported dividing issues into two moral domains: personal-sexual and violent-dishonest issues. After dividing issues by moral domain, GPT's estimates were still better in high than low-income countries, suggesting there may indeed be some cultural bias in the model. Yet, there was an even greater difference in prediction success between domains: in both high-income and low-income countries, GPT performed well in the personal-sexual domain but very poorly in the violent-dishonest domain.

The difference in performance between the two moral domains raises the question of what defines GPT's one-dimensional moral worldview. To explore this issue further, we probed GPT for country-level estimates of liberalism/conservatism for all 63 countries in the WVS dataset. These estimates were nearly perfectly correlated with GPT's public opinion estimates, suggesting that liberalism versus conservatism constitutes the core dimension in GPT's moral worldview. While the countries' degree of liberalism seems to predict moral opinions in the personal-sexual domain effectively, its relevance diminishes when applied to the violent-dishonest domain. This observation corresponds with prior research indicating that the autonomy–embeddedness dimension explains cultural variance in attitudes toward personal-sexual but not dishonest-illegal moral issues (Vauclair & Fischer, 2011). As Schwartz (2004) has noted, his autonomy–embeddedness dimension shares significant conceptual similarities with other dimensions that differentiate individual and collective concerns, among which the degree of conservatism/liberalism can be considered.

In summary, GPT's moral worldview suffers from a flaw as it flattens the moral landscape onto a single liberal–conservative dimension. Nonetheless, this dimension proves remarkably effective in explaining cultural variance in attitudes toward personal-sexual moral issues.

*Do* (arXiv:2103.11790). arXiv. https://doi.org/10.48550/arXiv.2103.11790

Schwartz, S. (2004). *Mapping and interpreting cultural differences around the world* (pp. 43–73). https://doi.org/10.1163/9789047412977_007

Singh, S., & Ramakrishnan, N. (2023). *Is ChatGPT Biased? A Review*. OSF. https://doi.org/10.31219/osf.io/9xkbu

Tao, Y., Viberg, O., Baker, R. S., & Kizilcec, R. F. (2023). *Auditing and Mitigating Cultural Bias in LLMs* (arXiv:2311.14096). arXiv. http://arxiv.org/abs/2311.14096

Vauclair, C.-M., & Fischer, R. (2011). Do cultural values predict individuals' moral attitudes? A cross-cultural multilevel approach. *European Journal of Social Psychology*, *41*(5), 645–657. https://doi.org/10.1002/ejsp.794

WVS/EVS. (2020, November 13). *Joint EVS/WVS 2017-2022 dataset*. European Values Study. https://europeanvaluesstudy.eu/methodology-data-documentation/survey-2017/joint-evs-wvs/